# Lattice displacements above $T_C$ in the layered manganite $La_{1.2}Sr_{1.8}Mn_2O_7$

D.N. Argyriou , H.N. Bordallo
*Manuel Lujan Jr. Neutron Scattering Center,*
*Los Alamos National Laboratory, Los Alamos, NM87545*

J.F. Mitchell, J.D. Jorgensen
*Material Science Division, Argonne National Laboratory, Argonne, IL 60439*

G.F. Strouse[*]
*Department of Chemistry, University of California Santa Barbara, Santa Barbara, CA 93106*

Neutron diffraction data presented in this paper demonstrates the relevance of lattice displacement above $T_C$, in our understanding of the evolution of the crystal structure with temperature in the layered CMR manganite $La_{1.2}Sr_{1.8}Mn_2O_7$. The anomalous temperature behavior of thermal diffuse scattering (TDS) in $La_{1.2}Sr_{1.8}Mn_2O_7$ strongly suggests that it arises from lattice displacements and correlates directly with anomalies in the displacement parameters of the O- and Mn-atoms and Mn-O bond lengths. From our measurements, the insulator - metal transition can be described as a transition from a high temperature state with disordered Mn-O bond lengths to a low temperature state with a more uniform distribution on Mn-O bonds. These observations are in agreement with polaronic charge transport above $T_C$ in the perovskite manganites; as electron hopping is responsible for bond disorder above $T_C$ , below the transition where $e_g$ carries are delocalized, any lattice displacements are uniformly averaged.

The strong coupling between magnetism and charge transport in the manganite perovskites has been known since the 1950's.[1, 2] Their recent rediscovery has centered on the observation of phenomena such magnetotransport[3] and magnetostructural effects[4] that highlight the interplay between spin, charge and lattice degrees of freedom, in these materials. Discussions on the physical origin of colossal magnetoresistance (CMR) in the manganite perovskites originally focused on the double exchange (DE) of $e_g$ carriers between ferromagnetically coupled $Mn^{3+}$ and $Mn^{4+}$ ions[5]. However, recent theoretical work suggests that the DE model alone does not quantitatively explain the CMR effect on $La_{1-x}M_xMnO_3$.[6, 7] Rather, strong electron-phonon coupling arising from local Jahn-Teller (JT) distortions of $Mn^{3+}$ ions due to the slow hopping of $e_g$ carriers may explain the magnetotransport and magnetostructural phenomena observed in these materials. Isotopic labeling experiments have highlighted the importance of polaronic degrees of freedom on the electronic and magnetic transition in the perovskite manganites.[8] It has become increasingly evident that the correlation between local structural changes and polaron formation can provide an adequate description for the anomalies observed in many experiments. For example unusual temperature dependence of displacement parameters (Debye-Waller factors) and lattice parameters,[9] large frequency shifts of the internal infrared modes,[10] variations in the local structure as seen by pair distribution function (PDF) analysis[11] and EXAFS.[12]

While the work on CMR materials has concentrated on the 3D perovskite manganites, the discovery of the layered compound $La_{2-2x}Sr_{1+2x}Mn_2O_7$,[13] as another class of CMR oxides, provides a rich opportunity to explore the interplay between spin, charge and lattice in reduced dimensions. These compounds crystallize in the body-centered-tetragonal structure I4/mmm ($D_{4h}^{17}$); which is built up of perovskite bilayers of corner-linked slightly distorted $MnO_6$ octahedra forming infinite sheets, separated by a (La,Sr)O layer along the *c*-axis.[14] An important structural feature of these materials is that the delocalization of charges results in an enhancement of the distortion of the $MnO_6$ octahedron in the metallic regime, a behavior that is in sharp contrast to the perovskite manganites. This enhancement in the $MnO_6$ octahedral distortion is

Permanent Address, Materials Science Division, Argonne National Laboratory, Argonne, IL, 60439
Permanent Address, Intense Pulsed Neutron Source, Argonne National Laboratory, Argonne, IL, 60439

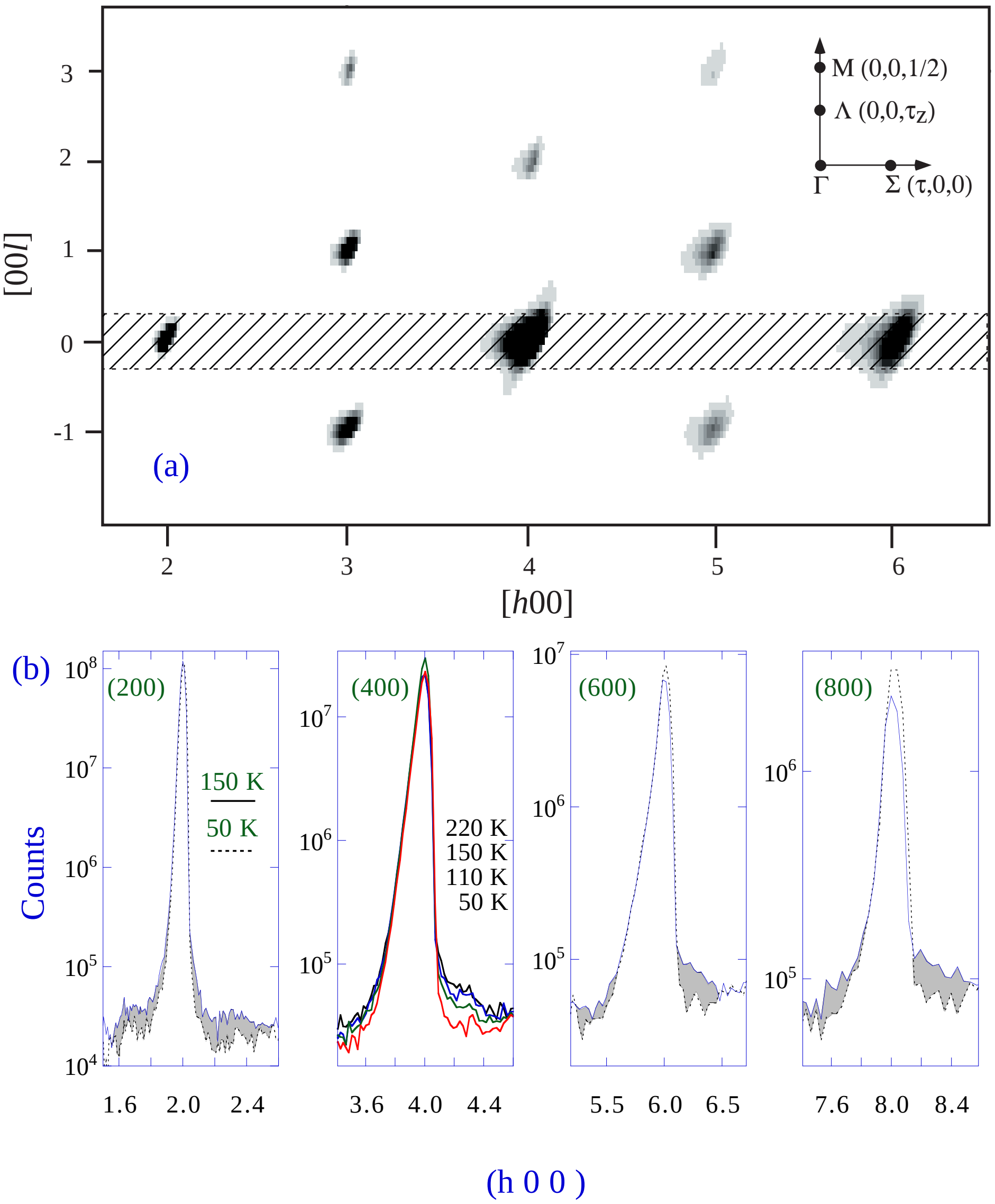

**Fig. 1:** (a) A section of [010] zone measured on the SCD spectrometer from a $La_{1.2}Sr_{1.8}Mn_2O_7$ single crystal at 150 K. The shaded area shows the integration performed along the <ξ 0 0> zone axis shown in figure 1(b). Some points on the first Brilloiun zone are shown in the inset. (b) Single crystal diffraction intensities for the (200), (600) and (800) reflections measured at 50 (dashed line) and 150 K (continuous line). The temperature dependence of the intensity of the diffuse scattering is shown for the (400) reflection. The shaded regions indicate the observed TDS scattering .

driven by an abrupt contraction of the equatorial Mn-O(3) bond and expansion of the Mn-O(2) bond at $T_C$ in the x=0.4 material.[14] However this structural responce to charge delocolization is not incompatible with a picture of lattice polarons. Recent Raman results[15] from a single crystal of $La_{1.2}Mn_{1.8}Mn_2O_7$ and PDF analysis of neutron diffraction data from a ceramic sample of $La_{1.4}Sr_{1.8}Mn_2O_7$ [16] suggest a significant electron-phonon coupling may accompany both the electronic and the magnetic transitions. To date the role of polarons in the physics of layered CMR manganites has not been explored fully and requires further study.

In this paper we report both time of flight (ToF) single crystal and powder neutron diffraction measurements from crystalline samples of the layered manganite $La_{1.2}Sr_{1.8}Mn_2O_7$ (x = 0.4). Apart from magnetic diffuse scattering at low **Q** arising from Mn spin correlations [17] significant amounts of thermal diffuse scattering (TDS) around reciprocal lattice points are observed. In the layered CMR $La_{1.2}Sr_{1.8}Mn_2O_7$ the temperature dependence of the TDS intensity is peculiar. From our data it is clear that the unusual temperature dependence of the TDS intensity is similar to the temperature dependence of displacement parameters (Debye-Waller factor) for the Mn- and O(3)-atoms, which abruptly decreases at $T_C$ as well. This phenomenon correlates strongly with significant lattice effects and charge delocalization measured in this sample and suggests that lattice displacements or disorder above $T_C$ is associated with the hopping of $e_g$ carriers, while such disorder relaxes below $T_C$ as $e_g$ carries become delocalized.[6, 7]

$La_{1.2}Sr_{1.8}Mn_2O_7$ single crystals were melt-grown in flowing 100% $O_2$ (balance Ar) in a floating zone optical image furnace (NEC SC-M15HD). Sample characterization in terms of transport, magnetization and neutron diffraction has been reported elsewhere.[14] From previous work, it is known that this sample exhibits a coupled electronic and ferromagnetic transitions at $T_C$~120 K. Single crystal ToF measurements have been performed as a function of temperature (20-300K) using the Single Crystal Diffractometer (SCD) at the Manuel Lujan Jr. Neutron Scattering Center at the Los Alamos National Laboratory. Displacement parameters U were obtained from Rietveld analysis of neutron powder diffraction measurements from a pulverized sample from the same boule using the Special Environment Powder Diffractometer (SEPD), between 20-550K at the Intense Pulsed Neutron Source at the Argonne National Laboratory.

ToF single crystal neutron diffraction provides the opportunity to measure large volumes of reciprocal space for both long range atomic ordering (Bragg reflections) as well as diffuse scattering that reflects structural features on a local scale. In Fig. 1(a) we show a measurement of reciprocal space perpendicular to the [010] zone-axis, while the inset shows various points of the first Brillouin zone for I4/mmm $-\Gamma_q^v(a), \tau < \tau_z$. Close inspection of the diffraction data, reveals a significant amount of diffuse scattering around Bragg reflections along the <ξ 0 0> direction. This is clearly seen in fig. 1(b) where we have integrated the three dimensional diffraction data along the <ξ 0 0> direction, over the shaded portion as shown in fig. 1(a). Since the data have been obtained using pulsed neutrons, the additional diffuse scattering is best observed before the fast rising, leading edge of the Bragg peak, than the slower trailing edge, although some evidence of diffuse scattering is seen there also. We find that this diffuse scattering is temperature dependent as shown in fig. 1(b) for the (400) reflection, and therefore may originate from a dynamical response.

To analyze this response, we have taken into account that in a real solid the atomic motion gives rise to scattering intensity that is distributed continuously (but not uniformly) throughout reciprocal space.[18] It is well known that the existence of some dynamical disorder in a crystal structure is revealed by the presence of a quasi-elastic signal in Raman spectroscopy or in inelastic neutron scattering, as found in ferroelectric systems[19] and by diffuse scattering around a Bragg peak.[20] A typical example are soft materials such as Pb, where diffuse scattering is observed around Bragg reflections due to the inelastic scattering of neutrons from acoustic phonon modes[18] The origin of the TDS lies in the $1/\omega_j^2(\mathbf{q})$ dependence of the first order scattering by lattice vibrations. Because the frequency $\omega_j(\mathbf{q})$ of the acoustic modes is proportional to $\mathbf{q}$ for small $\mathbf{q}$, (where $\mathbf{q}$ is point in the first Brillouin zone away from a reciprocal lattice vector $\mathbf{H}$) the TDS near the Bragg peak arising from these modes varies approximately as $1/\mathbf{q}^2$. The TDS from optic modes will tend to remain roughly constant across the Bragg peak.[18] Assuming that only the acoustic modes of vibration are responsible for TDS and that these modes propagate with the same velocity $v_s$ (with $v_s = \omega_j(\mathbf{q})/\mathbf{q}$) in all directions, close to a Bragg reflection $\mathbf{q} \to 0$, the intensity distribution $I_H(j\mathbf{q})$ in the first Brillouin zone is given by:[18]

$$I_H(j\mathbf{q}) \approx \frac{k_B T}{\rho v_s^2 \mathbf{q}^2} Q^2 \left| G_H(j\mathbf{q})^2 \right| \qquad \text{(I)}$$

where, ρ is the density of the crystal, $v_s$ is the velocity of sound in the material, Q is the magnitude of the scattering vector ($\mathbf{Q}=2\pi\mathbf{H}+\mathbf{q}$), and $G_H(j\mathbf{q})$ is the structure factor for first order (one-phonon) scattering that includes the structure factor for Bragg scattering (F($\mathbf{H}$)), the displacement parameter term (Debye-Waller factor) and a phonon term.

The dependence of the diffuse scattering intensity $I_H(j\mathbf{q})$ as a function of $\mathbf{Q}^2$, for $\mathbf{q}$ along <ξ 0 0>, with $\mathbf{q} \to 0$, for temperatures above $T_C$, is given in Figure 2(a). The observed linear dependence is in agreement with expression (I) and indicates that the diffuse scattering observed on $La_{1.2}Sr_{1.8}Mn_2O_7$ is similar to that expected from TDS. On the other hand, the intensity of the TDS is expected to vary linearly with temperature as again indicated by equation (I). As shown in fig. 2(b), the evolution of the TDS intensity as a function of temperature is quite unsuall, showing the expected linear behavior above $T_C$, and an abrupt decrease below $T_C$. This is in contrast to equation (I), suggesting that it may arise from a dynamic response as opposed to less temperature dependent diffuse scattering caused by defects in the crystal lattice. Furthermore, the anomaly seen in the temperature dependence of the TDS intensity correlates well with both the lattice effects observed in this material and the anomalies in the temperature dependence of the displacement parameters (Debye-Waller factors).

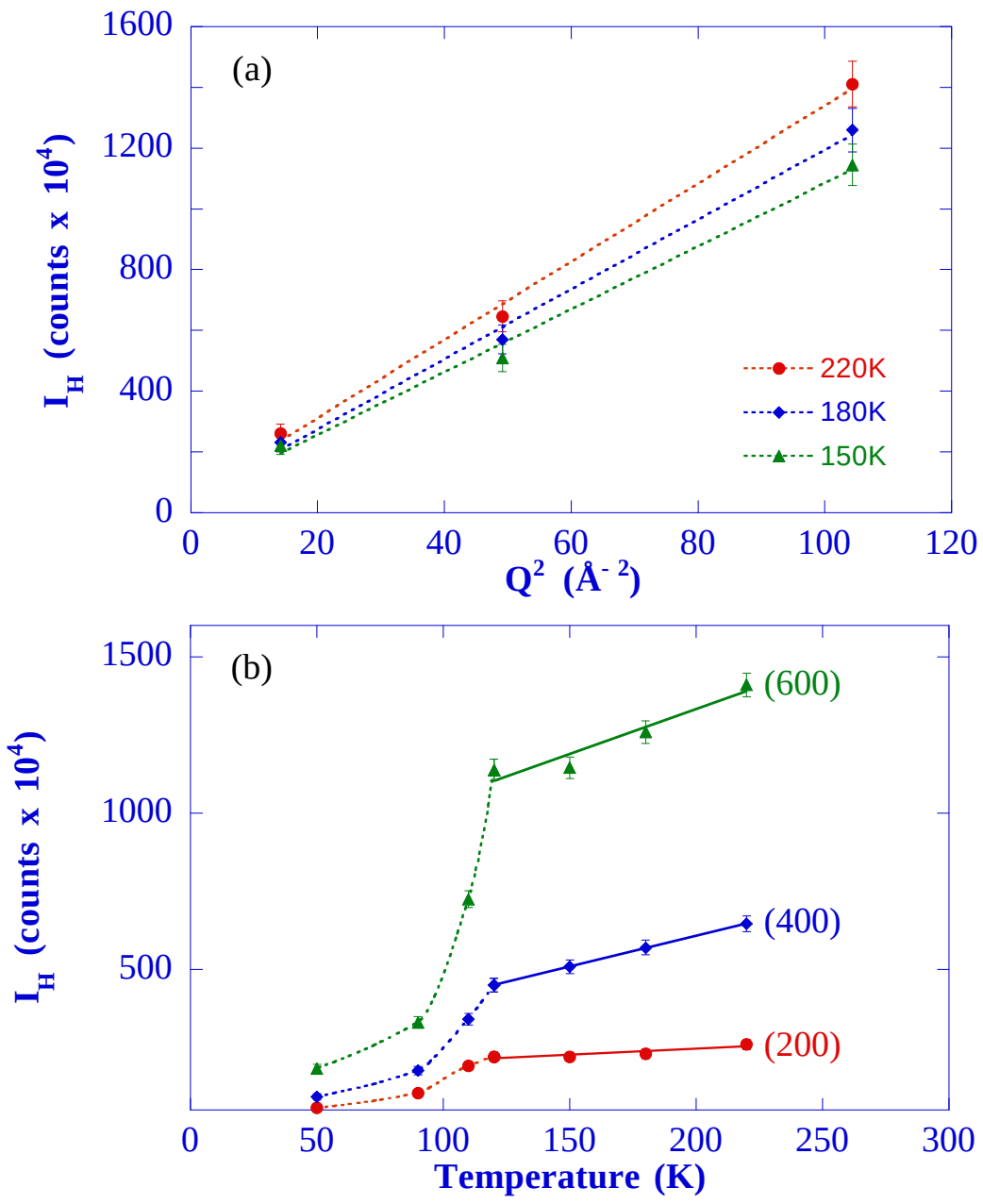

**Fig. 2:** Dependence of the TDS intensity (a) as a function of $\mathbf{Q}^2$ for T > 100 K, and (b) as function of temperature. We note the abrupt decrease of the intensity at $T_c$.

In fig. 3 we show the temperature dependence of the displacement parameters (U) for the three oxygen and Mn atoms, as well as the temperature dependence of the three Mn-O bonds, as determined from previous Rietveld analysis of powder diffraction data.[14] As seen in Figures 3(c) and 3(d), accompanying charge delocalization and magnetic ordering there are pronounced lattice effects resulting from changes in the Mn-O(2) and Mn-O(3) bond lengths. In addition to these changes notable variations in the displacement parameters for the Mn- and O-atoms are also observed.

At high temperatures the displacement parameter of all atoms varies linearly with temperature as expected.[18] However close to $T_C$, the temperature dependence of the TDS intensity, the Mn-O bonds and the displacement parameters show a very similar anomaly. This is especially the case for the Mn displacement parameter which shows a linear behavior from ~500K to 300 K, and then saturates to a constant value between $T_C$ and 300 K (Fig. 3(a)). This same behavior is observed for the Mn-O(3) bond length (Fig. 3(d)). At $T_C$ the displacement parameters for both the Mn- and O(3)-atoms show at abrupt decrease, ($\Delta U_{T_C}$ ~0.0016 $^{2}$ for O(3) and 0.001 $^{2}$ for the Mn-atom) accompanying the contraction of the Mn-O(3) bond. A smaller decrease of the displacement parameter is seen for the apical O(2) oxygen, despite the significant expansion of the Mn-O(2) at $T_C$ (Fig. 3(c)), while no significant anomaly is observed for either Mn-O(1) bond or the O(1) displacement parameter (Fig. 3(b)).

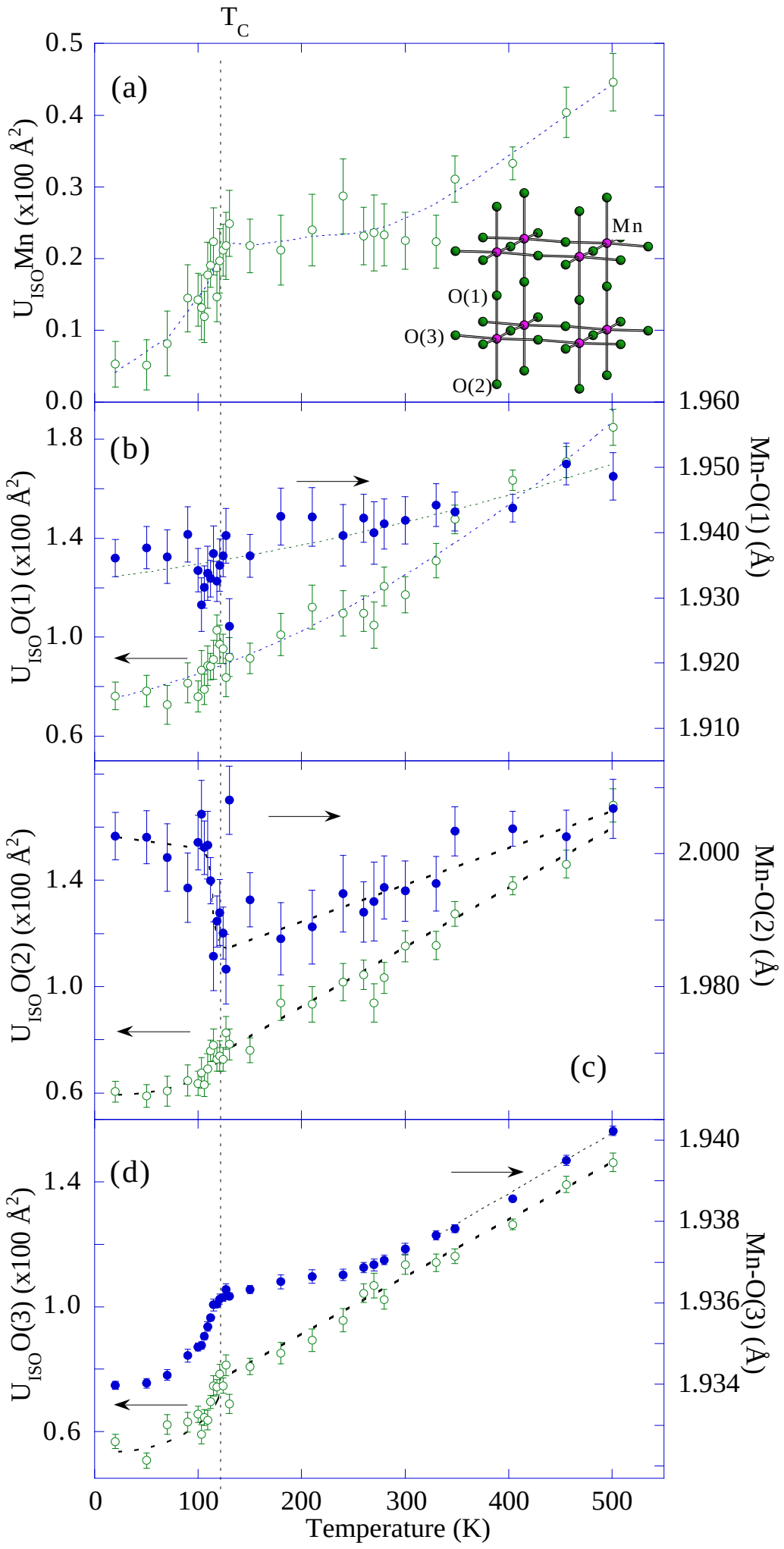

**Fig. 3**: Isotropic displacement parameters for the (a) Mn, (b) O(1), (c) O(2) and (d) O(3) atoms as a function of temperature. The small inset in fig. 3(a) depicts a single perovskite bilayer in the $La_{1.2}Sr_{1.8}Mn_2O_7$ structure with the labeling of the three oxygen sites. The temperature dependence of the Mn-O(1), Mn-O(2) and Mn-O(3) bonds are also shown in figures (b), (c) and (d) respectively.

Our measurements of both the Mn-O(3) bond and the displacements parameters of the Mn and O(3) atoms are linear at high temperature but exhibit a plateau between 300K and $T_C$ as shown in fig (3). We suggest that this plateau is caused by polaronic degrees of freedom inthis material that gives rise to TDS like scattering. Above 300K were the expected linear temperature dependence of these parameters resumes may suggest that the amplitude of normal thermal displacements becomes greater than the amplitude of Mn and O displacements that are coupled via the Jahn-Teller effect, to hopping $e_g$ charge carriers.

The relationship between the displacement parameter and the TDS scatering lies in that the displacement parameter reflects the spatial distribution of an atom in real space, which in turn attenuates the intensity of a Bragg reflection as exp(-1/2 $Q^2 \bullet U(k)$). The observation of an abrupt reduction of the TDS intensity at $T_C$ coinciding with a decrease of the displacement parameter suggests that the TDS intensity represents a 1-exp(-1/2 $Q^2 \bullet U(k)$) or more simply the intensity removed from a Bragg reflection due to disorder. In this case, the amplitude of the local displacements with respect to the long-range structure is entirely contained in the displacement-parameters.[18]

Our measurements suggest that polaronic degrees of freedom are also relevant in the layered CMR manganites $La_{1.2}Sr_{1.8}Mn_2O_7$. We find that the structural response to the coupled electronic and magnetic transitions in $La_{1.2}Sr_{1.8}Mn_2O_7$ results from lattice displacements, from a state characterized by a broad distribution of Mn-O bond lengths above $T_C$, to a state with a sharper Mn-O bond length distribution below $T_C$. Indeed this observation agrees with the prediction of Millis *et al.*[6] where a RMS oxygen displacements of >0.1 are predicted above $T_C$ becoming smaller below $T_C$. From our measurements the RMS displacements of the O(3) atom decreases from 0.089(2) at $T_C$ to 0.070(2) below $T_C$ and for the Mn atom from 0.046(2) to 0.023(3) respectively. Further the observation of the TDS scattering and its temperature dependence suggest that the observed lattice displacements may be dynamic in nature in agreement with the notion of the slow hopping of $e_g$ carrier and the associated JT distortion of $Mn^{3+}$ (and relaxation to $Mn^{4+}$) in the lattice.[6,7]

We acknowledge R. Osborn and J. Eckert for helpful discussions. The Manuel Lujan Jr. Neutron Scattering Center is a national user facility funded by US DOE-BES, under contract W-7405-ENG-36 (HNB, DNA) with the University of California. Argonne National Laboratory is funded by US DOE under contract W-31-109-ENG-38 (JFM, JDJ). The work performed by DNA, GSF and HNB was also supported by UCDRD grant STB-UC:97-240.